\documentclass[10pt,preprint,a4paper]{aastex}
\usepackage{amsmath}                
\usepackage{amsfonts}               
\usepackage{amssymb}                
\usepackage{epsfig}                 

\newcommand{\cm}{{~\rm cm}}
\newcommand{\km}{{~\rm km}}

\newcommand{\msec}{{~\rm ms}}
\newcommand{\g}{{~\rm g}}
\newcommand{\G}{{~\rm G}}

\newcommand{\erg}{{~\rm erg}}

\begin{document}

\title{The magnetar model of the superluminous supernova GAIA16apd and the explosion jet feedback mechanism (JFM)}

\author{Noam Soker\altaffilmark{1}}

\altaffiltext{1}{Department of Physics, Technion -- Israel Institute of Technology, Haifa
32000, Israel; soker@physics.technion.ac.il}

\begin{abstract}
Under the assumption that jets explode core collapse supernovae in a negative jet feedback mechanism (JFM), I show that rapidly rotating neutron stars are likely to be formed when the explosion is very energetic. Under the assumption that an accretion disk or an accretion belt around the just-formed neutron star launch jets and that the accreted gas spins-up the just-formed neutron star, I derive a crude relation between the energy that is stored in the spinning neutron star and the explosion energy. This relation reads
 $(E_{\rm NS-spin}/E_{\rm exp}) \approx E_{\rm exp}/10^{52} \erg$.
It shows that within the frame of the JFM explosion model of  core collapse supernovae, spinning neutron stars, such as magnetars, might have significant energy in super-energetic explosions. The existence of magnetars, if confirmed, such as in the recent super-energetic supernova GAIA16apd, further supports the call for a paradigm shift from neutrino-driven to jet-driven core-collapse supernova mechanisms.
\end{abstract}

\section{INTRODUCTION}
\label{sec:intro}

The two contesting explosion mechanisms of core collapse supernovae (CCSNe) to utilize the gravitational energy that is released during the formation process of the neutron star (NS; or black hole) are the delayed neutrino mechanism (e.g., \citealt{Bruennetal2016, Jankaetal2016, Muller2016, Burrowsetal2017}; for recent papers), and the jet feedback mechanism (JFM; e.g., \citealt{Papishetal2016, Gilkisetal2016}, and \citealt{Soker2016Rev} for a recent review). The collapse-induced thermonuclear explosion (CITE) mechanism \citep{Burbidgeetal1957, KushnirKatz2014} is based on nuclear energy, and it requires a large amount of angular momentum \citep{Kushnir2015a}. This implies that an accretion disk is formed around the newly born NS (\citealt{Gilkisetal2016};  confirmed by \citealt{BlumKushnir2016}), and jets that are expected to be launched by the accretion disk dwarf the energy of the thermonuclear reactions (\citealt{Gilkisetal2016}).

The delayed neutrino mechanism encounters some difficulties and it is not clear whether it can explain even a small fraction of CCSNe (e.g., \citealt{Papishetal2015, Kushnir2015b}).
In any case, even the most optimistic studies show that the delayed neutrino mechanism cannot explain CCSN explosion energies of $E_{\rm exp} \ga 2 \times 10^{51} \erg$ (e.g., \citealt{Fryer2006, Fryeretal2012, Sukhboldetal2016, SukhboldWoosley2016}; by explosion energy I refer to the kinetic energy and the radiated energy of the CCSN).
For that, the quest for the powering mechanism of super-energetic SNe (SESNe; $E_{\rm exp} \ga 10^{52} \erg$ ) is a hot topic (e.g., \citealt{GalYam2012, Moriyaetal2015, Wangetal2016, Arcavietal2016, Sorokinaetal2016, LiuModjaz2017}).

Many studies attribute the extra energy of very luminous CCSNe, termed superluminous SNe, or of SESNe to magnetars, i.e., strongly-magnetized, rapidly-rotating NS (e.g., \citealt{KasenBildsten2010, Woosley2010, Metzgeretal2015}).
In a recent paper \citep{Soker2016Mag} I show that under reasonable assumptions the formation process of a magnetar will be accompanied by the launching of energetic jets, and that the jets can carry an amount of energy that surpasses the energy that is stored in the newly born magnetar. This raises the possibility that some superluminous CCSNe are powered by late jets as part of the JFM  (e.g., \citealt{Gilkisetal2016}), rather than by a magnetar.

A recent superluminous CCSN for which the magnetar model was applied is GAIA16apd.
\cite{Yanetal2017} estimate the ejected mass of GAIA16apd to be $M_{\rm ej} \simeq 12 M_\odot$, and for an opacity of $\kappa = 0.1 \cm^2 \g^{-1}$ derived an explosion energy of $E_{\rm exp} > 2 \times 10^{52} \erg$.
\cite{Nicholletal2017} present a magnetar explanation for the superluminosity. In their model the ejected mass is $M_{\rm ej} = 4(0.2 \cm^2 \g^{-1}/\kappa) M_\odot$, the spin period of the pulsar is $P=2 \msec$, and its magnetic field is $B=2 \times 10^{14} \G$.
\cite{Kangasetal2016} has a consistent magnetar model, within the uncertainties, with $M_{\rm ej} = 7(0.2 \cm^2 \g^{-1}/\kappa) M_\odot$, and explosion energy of $M_{\rm ej} = 1\times 10^{52} (0.2 \cm^2 \g^{-1}/\kappa) \erg$, and for the magnetar they deduce $P=1.9 ^{+0.3}_{-0.2} \msec$, and  $B=2.1^{+0.5}_{-0.2} \times 10^{14} \G$.

Not only such a superluminous CCSN requires a long lasting central engine, such as jets or a magnetar, but the explosion itself most likely is driven by jets.
The possibility that jets play a role in the explosion process of massive stars has been mentioned in the literature over the years. Three examples, among many others, are the recent paper by \cite{Mauerhanetal2017} who discuss the axisymmetrical explosion of SN~2013EJ, the polarimetric observations of SN~2015bn that indicate an elongated morphology \citep{Inserraetal2016}, and the presence of jets in CCSN remnants (e.g., \citealt{Lopezetal2014, Milisavljevic2013}). As well, many studies simulated jets in CCSNe (e.g., \citealt{BrombergTchekhovskoy2016}). However, the majority of these papers refer to jets only in rare CCSNe and/or as an additional component to the explosion mechanism that was assumed to be driven by neutrinos. I take the view that in all these cases, and in the majority of (or even all) regular CCSNe, jets explode the star rather than neutrinos, and that the jets operate in a negative feedback mechanism (see \citealt{Soker2016Rev} for a review). For example, under this view the association of gamma ray bursts with type Ic supernovae (e.g., \citealt{Canoetal2016, Modjazetal2016}) is a result of the explosion being driven by jets.

In the present study I explore the implications of magnetar formation in the context of the explosion JFM. When pre-collapse rapidly rotating cores collapse, they form an accretion disk around the newly born NS or black hole (e.g., \citealt{Gilkis2017}), and jets are likely to be launched (e.g., \citealt{Nishimura2015}). I take this into account in evaluating the connection between a newly born magnetar and the JFM.
I do not claim that magnetar is the only possible powering mechanism of GAIA16apd or similar SESNe, as jets might also account for extended operation of the central engine. I simply reveal the implications of the existence of a magnetar.

\section{ANGULAR MOMENTUM AND ENERGY}
\label{sec:angular}

\cite{Nicholletal2017} use the magnetar model presented by \cite{Inserraetal2013} who take the angle between the magnetic axis and spin axis to be $45^\circ$. If the angle is smaller, then a more rapid rotation is required to have the same cooling time. So one should bear in mind the uncertainties in the values of spin period and energy of the magnetar model.

As discussed in section \ref{sec:intro}, I think that the JFM is the only viable model to account for the explosion of GAIA16apd. I now discuss the implication of accretion of mass and angular momentum onto the just-formed (newly born) NS.
I assume that after the shock bounces from the just-formed NS, the NS does not rotate rapidly.
I consider then that a mass $M_{\rm acc}$ is accreted onto the newly born NS, and launches jets.
The accreted gas can form an accretion disk or an accretion belt \citep{SchreierSoker2016} .

The energy that is accreted through the accretion disk or accretion belt and transferred to the jets is taken to be
\begin{equation}
E_{\rm exp} \simeq E_{\rm jets} = \xi \frac{G M_{\rm NS} M_{\rm acc}}{2 R_{\rm NS}}
= 2.6 \times 10^{52} \xi     
\left(  \frac{M_{\rm NS}}{1.5 M_\odot} \right)
\left(  \frac{R_{\rm NS}}{15 \km } \right)^{-1}
\left(  \frac{M_{\rm acc}}{0.2 M_\odot} \right) \erg ,
\label{eq:exp1}
\end{equation}
where $M_{\rm NS}$ and $R_{\rm NS}$ are the neutron star mass and radius, respectively, and $\xi$ is the efficiency by which gravitational energy is channelled to the kinetic energy of the jets.
Strictly speaking, the binding energy of the ejected mass should be removed from the energy in the jets to obtain the explosion energy. But for the discussed type of superluminous CCSNe the explosion energy is much larger than the binding energy.
{{{ Note that I defined $\xi$ above with a factor of 2 in the denominator. I will later scale with $\xi=0.4$, that implies that the jets carried 20 per cents of  
$G M_{\rm NS} M_{\rm acc}/R_{\rm NS}$.  }}}

I take the specific angular momentum of the accreted mass, $j_{\rm acc} = J_{\rm acc}/M_{\rm acc}$, where $J_{\rm acc}$ is the total angular momentum of the accreted mass, to be about equal to that of the critical value (about Keplerian orbit) on the surface of the NS, i.e., 
$j_{\rm acc} \simeq ( {G M_{\rm NS}} {R_{\rm NS}} )^{1/2}$. I further assume that the final angular momentum of the NS is parameterized by $J_{\rm NS}= \beta J_{\rm acc}$. This is a crude estimate for the following processes. (1) The mass might be accreted onto a larger radius than $R_{\rm NS}$, hence with a larger angular momentum, before the NS makes the contraction to its final radius. (2) The final angular momentum of the NS can be larger if the just-formed NS has initial angular momentum. (3) The final angular momentum of the NS can be smaller if different segments of the accreted mass have different directions of angular momentum. This is the expectation in the jittering-jets model.  Despite the crude derivation, I think that it does present the correct trend, and it is adequate for the derivation to follow.

Under the above assumptions the angular momentum of the NS is
\begin{equation}
J_{\rm NS} = \beta J_{\rm acc} \simeq
\beta M_{\rm acc}  \left( {G M_{\rm NS}} {R_{\rm NS}} \right)^{1/2}.
\label{eq:am1}
\end{equation}
The angular momentum of the spinning NS is $J_{\rm NS}=I_{\rm NS} \omega$, where $\omega$ is the angular velocity of the NS, and the moment of inertia of a NS is $I= \eta M_{\rm NS} R^2_{\rm NS}$ with $\eta \simeq 0.3$ \citep{Worleyetal2008}. From equation (\ref{eq:am1}) one can find the ratio of the angular velocity to the critical (maximum) value $\omega_c$ of the NS to be  
\begin{equation}
\frac {\omega}{\omega_c} \simeq \frac {\beta M_{\rm acc}}{\eta M_{\rm NS}}
= 0.44 \beta
\left(  \frac{M_{\rm acc}}{0.2 M_\odot} \right)
\left(  \frac{M_{\rm NS}}{1.5 M_\odot} \right)^{-1}
\left(  \frac{\eta}{0.3} \right)^{-1} .
\label{eq:omega1}
\end{equation}

Under the above assumption on the final angular momentum of the NS, the ratio of the energy of the spinning NS, $E_{\rm NS,spin}=(1/2)I_{\rm NS} \omega^2$, to the explosion energy, which is about the energy carried by the jets, is
\begin{equation}
\frac{E_{\rm NS,spin}}{E_{\rm exp}} \simeq \frac{\beta^2}{\eta \xi} \frac{M_{\rm acc}}{M_{\rm NS}},
\label{eq:Erot1}
\end{equation}
where equation (\ref{eq:exp1}) has been used for the explosion energy and equation (\ref{eq:omega1}) for the angular velocity. A more transparent expression for the present goals can be obtained if equation (\ref{eq:exp1}) is used to replace the accreted mass by the explosion energy
\begin{equation}
\frac{E_{\rm NS-spin}}{E_{\rm exp}} \simeq \frac{2 \beta^2}{\eta \xi^2} \frac{R_{\rm NS}}{G M^2_{\rm NS}} E_{\rm exp}
\simeq 1   \beta^2   
\left(  \frac{\eta}{0.3} \right)^{-1}
\left(  \frac{\xi}{0.4} \right)^{-2}
\left(  \frac{M_{\rm NS}}{1.5 M_\odot} \right)^{-2}
\left(  \frac{R_{\rm NS}}{15 \km } \right)
\left(  \frac{E_{\rm exp}}{10^{52} \erg} \right).
\label{eq:Erot2}
\end{equation}
The meaning of this ratio is that in the JFM for exploding massive stars, the energy stored in the spinning NS is expected to be significant in SESNe, but not in regular CCSNe.

The ratio of spinning energy to the explosion energy depends stronger even on the explosion energy. For slowly pre-collapse rotating core, the specific angular momentum of the accreted mass is more stochastic (the jittering jets explosion mechanism), and the final angular momentum of the NS is lower, e.g., $\beta < 1$, and even $\beta \ll 1$, in equation (\ref{eq:am1}). This is a case of typical CCSNe. When the pre-collapse core is rapidly rotating, the JFM is less efficient, and the explosion energy is expected to be larger \citep{Gilkisetal2016}. In the case of a rapidly rotating pre-collapse core, the just-formed NS is expected to have angular momentum in the same sense as the accreted mass that forms jets, and hence the value of $\beta$ can be somewhat larger than 1.

The main conclusion from the simple derivation presented in this section is that the JFM for exploding CCSNe can account for rapidly spinning NS in the case of energetic explosions, i.e., for superluminous CCSNe that are not powered by collision of the ejects with the circum-stellar matter, and for SESNe. In other words, a magnetar is a possible outcome of the JFM explosion of SESNe.

\section{SUMMARY}
\label{sec:Summary}

Under the assumption that jets explode super-energetic supernovae (SESNe) in a negative jet feedback mechanism (JFM), I derived a crude relation between the explosion energy and the energy of the rotating NS at the end of the accretion phase, i.e., few seconds after core-collapse. This relation that is presented in equation (\ref{eq:Erot2}), shows that within the frame of the JFM explosion model of CCSNe, spinning NS, such as magnetars, can have significant energy only in super-energetic explosions (SESNe).

In other words, a rapidly rotating NS is a natural outcome of the JFM in the case of energetic explosions. In that sense, the existence of magnetars, if confirmed, such as in the recent SESN GAIA16apd, supports the JFM explosion model of CCSNe. I do note that it is possible that some superluminous CCSNe are powered by long-lasting jets rather than by a magnetar (e.g., \citealt{Gilkisetal2016}).

{{{ The point to emphasize following the present study is that the supernova explosion mechanism itself cannot be disregarded when studying the magnetar powering mechanism. In many cases, and possibly in all cases, the delayed neutrino mechanism cannot account for CCSNe that form magnetars, and jets must be considered as the explosion mechanism. }}}

The JFM model for exploding CCSNe is not yet in a stage of a developed model, as there are no simulations that demonstrate its operation from core collapse to explosion in different cases. At this stage, different observations and theoretical arguments are accumulated in small increments to support its potential as a viable mechanism for exploding most CCSNe. The present paper is another step forward in supporting the JFM as the major process to explode massive stars.

This research was supported by the Israel Science Foundation.

\end{document}